\documentstyle[11pt,newpasp,twoside,epsf]{article}
\def\edcomment#1{\iffalse\marginpar{\raggedright\sl#1\/}\else\relax\fi}
\reversemarginpar

\begin{document}
\title{Long Term Profile Variability of Double-Peaked Emission Lines in AGNs}
\author{Karen T. Lewis, Michael Eracleous}
\affil{Department of Astronomy and Astrophysics, The Pennsylvania State University, 
525 Davey Laboratory, University Park, PA 16802, USA}
\author{Jules P. Halpern}
\affil{Department of Astronomy, Columbia University, 550 West 120th Street, New York, 
NY 10027, USA}
\author{Thaisa Storchi-Bergmann}
\affil{Instituto de Física, UFRGS, 91501-970 Porto Alegre, RS, Brazil}

\begin{abstract}
An increasing number of AGNs exhibit broad, double-peaked Balmer
emission lines, which are thought to arise from the outer regions of
the accretion disk which fuels the AGN. The line profiles are observed
to vary on a characteristic timescales of 5-10 years. The variability
is not a reverberation effect; it is a manifestation of physical
changes in the disk. Our group has monitored a set of 20 double-peaked
emitters for the past 8 years (longer for some objects). Here, we
characterize the variability of the double-peaked H$\alpha$ line
profiles in five objects from our sample.  By experimenting with
simple models, we find that disks with a single precessing spiral arm
are able to reproduce many of the variability trends that are seen in
the data.
\end{abstract}

\section{Introduction}
An increasing number of AGNs are known to exhibit broad,
double-peaked, Balmer emission lines.  Approximately 20\% of
Broad-Line Radio Galaxies (BLRGs) are double-peaked emitters
(Eracleous et al. 1994).  Recently, over 100 double-peaked emitters
were discovered in the SDSS (Strateva et al. 2003). Double-peaked
emission lines have also appeared in objects which did not previously
exhibit them, most notably in the LINER NGC 1097 (Storchi-Bergmann et
al. 1993).  The lines are thought to originate in the outer portions
of the accretion disk (R $\sim$ 100s -1000s r$_{g}$, where r$_{g} =
GM_{\bullet}/c^{2}$).  The {\it profiles} of the double-peaked
emission lines are observed to vary on timescales of 5-10 yrs. These
variations are {\it not} a result of reverberation, which occurs on
the much shorter light-crossing timescale. The profile variability is
a manifestation of {\it physical} changes in the outer
disk. Therefore, detailed studies of the double-peaked emitters are
important, not only because they are part of an intriguing (and
growing) class of objects, but also because they are an important tool
which can be used to test models for the outer accretion disk. They
may also provide clues for the origin of the photometric variability
of all AGNs.

A simple axisymmetric disk can reproduce many of the individual
double-peaked profiles, but cannot explain the variability or the
commonly observed profiles in which the red peak is stronger than the
blue peak. The simplest extensions are axisymmetric disks with
emissivity perturbations, disks with precessing spiral arms, or
precessing elliptical disks. These models predict variability over
several different physical timescales, the dynamical, thermal, and
sound crossing times, which are set by the black hole mass
(M$_{\bullet}$) and given by: $\tau_{dyn}\sim 6\; \rm
M_8\;\xi_{3}^{3/2}$ months; $\tau_{th}\sim\tau_{dyn}/\alpha$; and
$\tau_{s}\sim 70\;\rm M_{8}\;\xi_{3}\;\rm T_5^{-1/2}$ years, where
$\rm M_{8}=\rm M_{\bullet}/10^8\;{\rm M}_{\odot}$, $\xi_{3} = \rm
R/10^{3}\rm R_{\rm g}$, $\rm T_{5}=\rm T/10^{5}\;$K, and $\alpha(\sim
0.1)$ is the Shakura-Sunyaev viscosity parameter (Shakura \& Sunyaev
1973). Matter embedded in the disk will orbit on the dynamical
timescale, thermal instabilities will dissipate over a thermal
timescale, while density perturbations will precess on timescales from
a few $\tau_{dyn}$ -- few $\tau_{s}$. Any of the above families of
models predict variability timescales that are roughly consistent with
those observed. In order to observe profile variability over the
different possible timescales, our group has been monitoring a set of
20 double-peaked emitters over the last decade. About half of the
objects are observed two or three times a year; others are accessible
only once each year. For a subset of these objects, the available data
span a longer baseline.

\section{Characterizing the Profile Variability}
Finding an appropriate model which can self-consistently reproduce a
sequence of profiles is a time-consuming process. Even within a family
of models, the parameter space to explore is large. As a {\it first
step}, to aid us in selecting suitable models, we have begun
characterizing the profile variability of the objects in our sample by
reducing each profile to a set of easily measured quantities whose
variability patterns can be quickly compared with those of a set of
model profiles. These parameters are: the velocities of the red and
blue peaks; the blue-to-red peak flux ratio; the full widths at half
and quarter maximum (FWHM and FWQM); and the velocity shifts of the
FWHM and FWQM centroids. Thus, for a given family of models we can
efficiently select the regions of parameter space that should be used
for detailed modeling for each object in our group.

Here, we compare the observed profile variability with that expected
from a circular disk with a single, precessing, spiral
arm. Chakrabarti and Wiita (1994) first suggested that spiral shocks
in AGN accretion disks might explain the observed line profile
variability in Arp 102B and 3C 390.0. This is an attractive model to
test first for several reasons. (1) Spiral arms have been used very
successfully in detailed modeling of sequences of profiles by Gilbert
et al. (1999; 3C 390.3 and 3C 332) and Storchi-Bergman et al.~(2003;
NGC 1097). (2) Spiral arms are present in other astrophysical disks
(i.e. galaxies and some cataclysmic variables). (3) A spiral arm
provides a mechanism for removing angular momentum from the disk. If
the profile variability is caused by the precession of a spiral arm,
detailed modeling could give us important information about how the
matter in the outer accretion disk sheds its angular momentum -
something which is not yet well understood!

\begin{figure}
\plotone{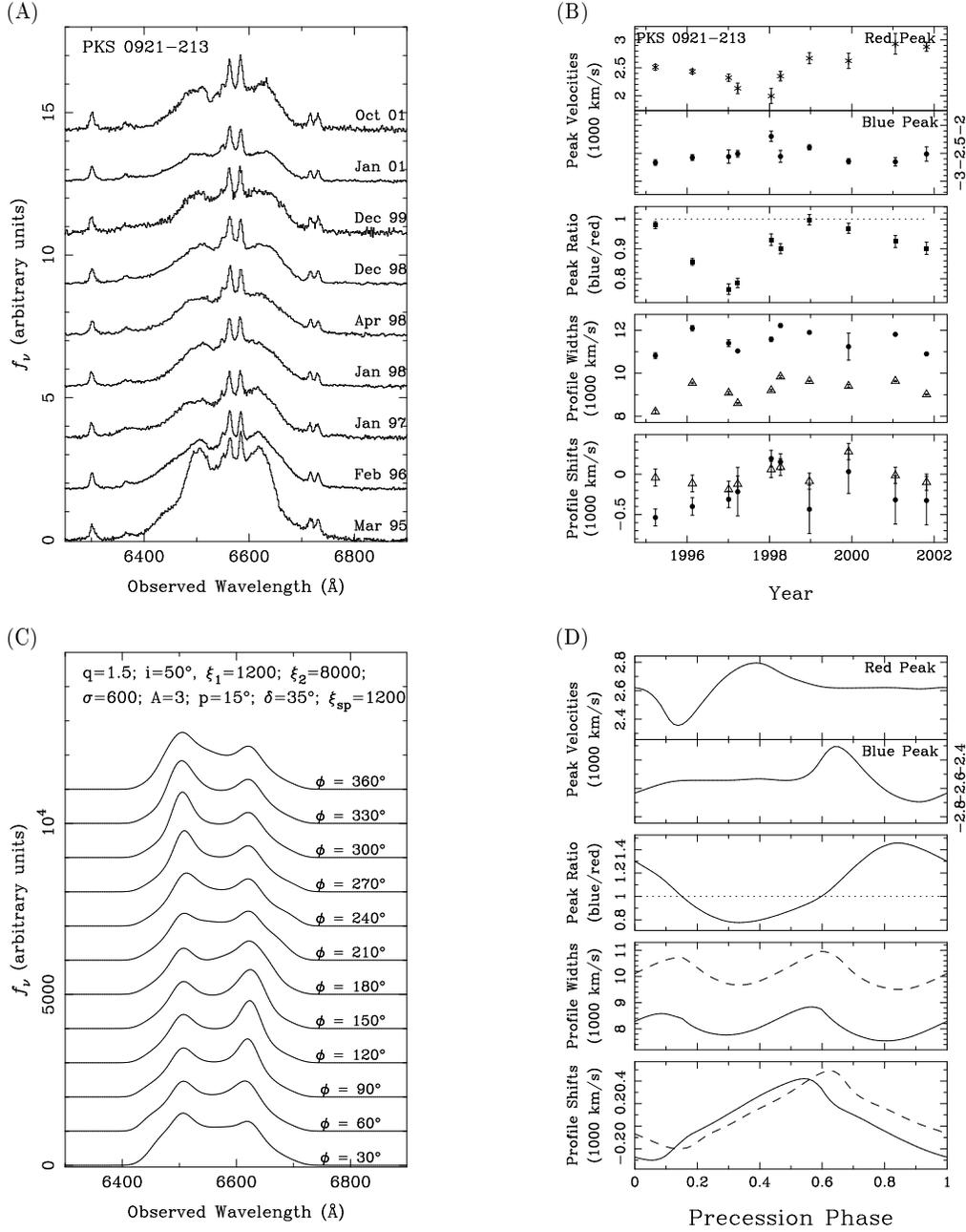}
\caption{ {\it PKS 0921--213 data} (A) Time sequence of
profiles for PKS 0921-213. The variation in the relative fluxes of the
red and blue peaks is quite clear. (B) Variability of the profile
properties with time. The FWHM is denoted with triangles, while the
FWQM is shown with filled circles. Observations taken within 1 month
of each other have been averaged together. The most striking variation
is the reversal in the blue to red peak flux ratio. {\it One-Armed
Spiral Model} (C) Example set of model profiles. The disk parameters
(see text for the definitions) have been tuned to fit the profile of
PKS 0921--213 while the spiral arm model parameters were chosen merely
to serve as an illustration. (D) The corresponding variability in the
model profile properties. The FWQM is shown with a dashed line.}
\end{figure}

The underlying axisymmetric disk has the following parameters: the
inner and outer radii, $\xi_{1}$ and $\xi_{2}$ (in units of
GM$_{\bullet}$/c$^{2}$); the inclination angle, $i$; a broadening
parameter, $\sigma$; and the slope of the axisymmetric emissivity law,
q ($\epsilon \propto \xi^{-q}$). The one-armed spiral is modeled as a
perturbation in the emissivity pattern, introducing the following
extra parameters: the amplitude, A, of the spiral arm (i.e. the
contrast relative to the underlying disk); the pitch angle, $p$; the
angular width of the arm, $\delta$; the inner radius of the arm,
$\xi_{sp}$; and the phase angle, $\phi$ (see Storchi-Bergmann et
al. 2003 for the full description). Profile sequences are produced by
varying $\phi$.

\section{Results}
Thus far, we have characterized the variability of five of the 20
objects in our sample (PKS 0921--213, CBS 74, PKS 1020--103, PKS
1739+184, and 3C 59). Each object is unique, but there are some common
trends which are exemplified in PKS 0921--213 (Fig.~1). (1) The most
striking variation is in the peak flux ratio. In some cases a complete
peak reversal is seen, {\it demanding} some kind of azimuthal
asymmetry in the disk. (2) The peak velocities show more moderate
modulations than the peak flux ratio. (3) The FWHM and FWQM vary
synchronously and by the same magnitude while the shifts of the FWHM
and FWQM centroids often vary at different times and by different
magnitudes.

The one-armed spiral models reproduce a wide range of variability
patterns.  But as a class, the spiral-arm models exhibit common
trends, some of which bear a striking resemblance to the data. An
example is shown in Figure~1. (1) Peak reversals are a ubiquitous feature
of the model profiles. (2) The blue and red peaks vary at different
times, but by the same magnitude. (3) The profile widths change
in-sync and by the same magnitude, as seen in the data. Although not
well demonstrated in Figure~1, in many models, the shifts of the FWHM
and FWQM centroids do not always vary synchronously.

This preliminary variability characterization suggests that a spiral
arm model is certainly capable of reproducing many of the variability
patterns that are observed. Additionally, this model has been used
quite successfully to model {\it sequences} of profiles. Therefore,
this model is likely to be an excellent candidate for detailed
modeling of these objects, and likely many more in our sample.

\end{document}